\pdfoutput=1

 \documentclass[final,5p,times,twocolumn]{elsarticle}

\usepackage{amsmath,graphicx,multirow,epsfig}
\usepackage[font=footnotesize]{caption}
\usepackage[font=footnotesize]{subcaption}
\usepackage{subcaption}
\usepackage{pgfplots}
\usepackage{pifont}
\usepackage{tikz}
\usepackage{array}

\journal{Signal Processing: Image Communication}

\begin{document}

\begin{frontmatter}



\title{Lossless Intra Coding in HEVC with 3-tap Filters}



\author[focal]{Saeed~R.~Alvar}
\ead{saeed.alvar@metu.edu.tr}

\author[focal]{Fatih~Kamisli}
\ead{kamisli@eee.metu.edu.tr}

\address[focal]{Department of Electrical and Electronics Engineering, Middle East Technical University, Turkey}

\begin{abstract}
This paper presents a pixel-by-pixel spatial prediction method for lossless intra coding within High Efficiency Video Coding (HEVC). A well-known previous pixel-by-pixel spatial prediction method uses only two neighboring pixels for prediction, based on the angular projection idea borrowed from block-based intra prediction in lossy coding. This paper explores a method which uses three neighboring pixels for prediction according to a two-dimensional correlation model, and the used neighbor pixels and prediction weights change depending on intra mode. To find the best prediction weights for each intra mode, a two-stage offline optimization algorithm is used  and a number of implementation aspects are discussed to simplify the proposed prediction method. The proposed method is implemented in the HEVC reference software and experimental results show that the explored 3-tap filtering method can achieve an average $11.34\%$ bitrate reduction over the default lossless intra coding in HEVC. The proposed method also decreases average decoding time by $12.7\%$ while it increases average encoding time by $9.7\%$.
\end{abstract}


\end{frontmatter}

\newcommand{\ry}{\rho_y}
\newcommand{\rx}{\rho_x}
\newcommand{\tu}{\ensuremath{u(i,j)}}
\newcommand{\tuq}[2]{\ensuremath{u(#1,#2)}}
\newcommand{\tuqh}[2]{\ensuremath{\hat{\tilde{u}}(#1,#2)}}

\section{Introduction}
\label{sec:intro}
Image\footnote{This research was supported by Grant 113E516 of T\"{u}bitak.} and video compression is performed either lossless or lossy. In lossless compression, the visual data is preserved perfectly. In lossy compression, some amount of degradation in the visual data is tolerated to achieve better  compression. Lossless and lossy compression are used in different applications. For example, in many multimedia applications such as TV broadcast or videoconferencing, lossy compression is used. In applications where perfect preservation of visual data is more important than bandwidth, such as film archiving or some medical applications, lossless compression is used.

Recently developed HEVC \cite{HEVC} or widely used H.264/AVC \cite{Luthra264} video coding standards support both lossy and lossless compression. Both lossy and lossless compression in these standards is achieved with a block-based approach. In lossy compression, a block of pixels are first predicted using pixels from a previously coded frame (inter prediction) or using pixels from previously coded regions of the current frame (intra prediction). The prediction is in many cases not accurate and as the next step, the block of prediction errors are computed and transformed to remove any remaining spatial redundancy. Finally, the transform coefficients are quantized and entropy coded together with other relevant side information such as prediction modes. 

In lossless compression, the transform and quantization steps are skipped and the prediction error block is directly entropy coded. However, the same block-based prediction methods are used. In the case of intra prediction, the block-based prediction approach becomes less efficient if the transform step is skipped. Therefore, a large number of research papers have proposed to modify  the block-based intra prediction approach used in lossless coding within H.264/AVC or HEVC.

There are two major approaches to improve the block-based intra prediction in lossless intra coding. In the first group of approaches, first the block-based intra prediction is performed and then the prediction error block is further processed with a second pixel-by-pixel prediction step \cite{sulivanDPCM,cross,Secondary,IRDPCM,ARDPCM}. In the second group of approaches, the block-based prediction approach is replaced directly with a pixel-by-pixel prediction approach \cite{SAP,DC2,Templatebased,ADSAP}. These approaches are discussed in more detail in the next section.

This paper explores an approach for lossless intra coding within HEVC which falls into the second group of approaches. In this approach, spatial prediction is performed in a pixel-by-pixel manner, however, the prediction equations are changed to improve the prediction performance and increase the compression efficiency. In particular, while many methods in the literature use two neighbor pixels for predicting the current pixel from a particular direction \cite{SAP} based on the angular projection idea borrowed from block-based intra prediction, this paper uses three pixels for prediction according to a two-dimensional correlation model. This can improve coding gains while merely increasing computational complexity. The used neighbor pixels and weights change depending on intra mode. To find the best weights for each intra mode, a two-stage offline optimization algorithm is utilized and a number of implementation aspects are discussed to simplify the proposed prediction method. Experimental results with the HEVC reference software show improved coding gains with respect to the default block-based prediction approach or other improved approaches in the literature.

The remainder of the paper is organized as follows. Section \ref{sec:pre_res} discusses related previous research, in particular, the default intra prediction method in HEVC and some major approaches for improving lossless intra coding. Section \ref{sec:pro_app} presents the proposed spatial prediction approach for lossless intra coding in HEVC, and discusses a number of related issues. Next, Section \ref{sec:exp_res} provides experimental results to compare the compression performance and computational complexity of the proposed approach with the default lossless intra coding in HEVC and other improved approaches. Finally Section \ref{sec:conc} concludes the paper.

\section{Previous Research }
\label{sec:pre_res}
This section discusses previous research closely related to this paper. First, a brief overview the block-based lossless intra coding in HEVC is given. Next, major approaches for improving the default lossless intra coding in HEVC are discussed, by categorizing them into three groups.

\subsection{Lossless intra coding in HEVC}
\label{ssec:loss_hevc}
Lossless intra coding in HEVC is achieved by retaining prediction and entropy coding and skipping transform, quantization, and in-loop filters \cite{HEVC}.To exploit spatial redundancy efficiently, block-based intra prediction in HEVC supports 35 modes for all Prediction Units (PU) with different sizes ranging from 4x4 to 64x64. Prediction modes 0 and 1 are called planar and DC modes, respectively and the remaining 33 modes are angular modes. While the planar and DC modes provide methods for effectively predicting image blocks with smooth or gradually changing content, the angular modes can effectively predict image blocks with directional structure.

In DC mode, each block pixel is predicted with the same value, which is obtained by averaging the neighbor pixels immediately to left and to the above of the block to be predicted. In planar mode, the block is predicted according to a plane model, which can provide a gradually changing prediction block (see \cite{IntraHEVC} for details).

\begin{figure}[tb]
  \centering
  \centerline{\includegraphics[scale=0.65, trim = 5.1cm 21.0cm 4.9cm 2.0cm, clip=true]{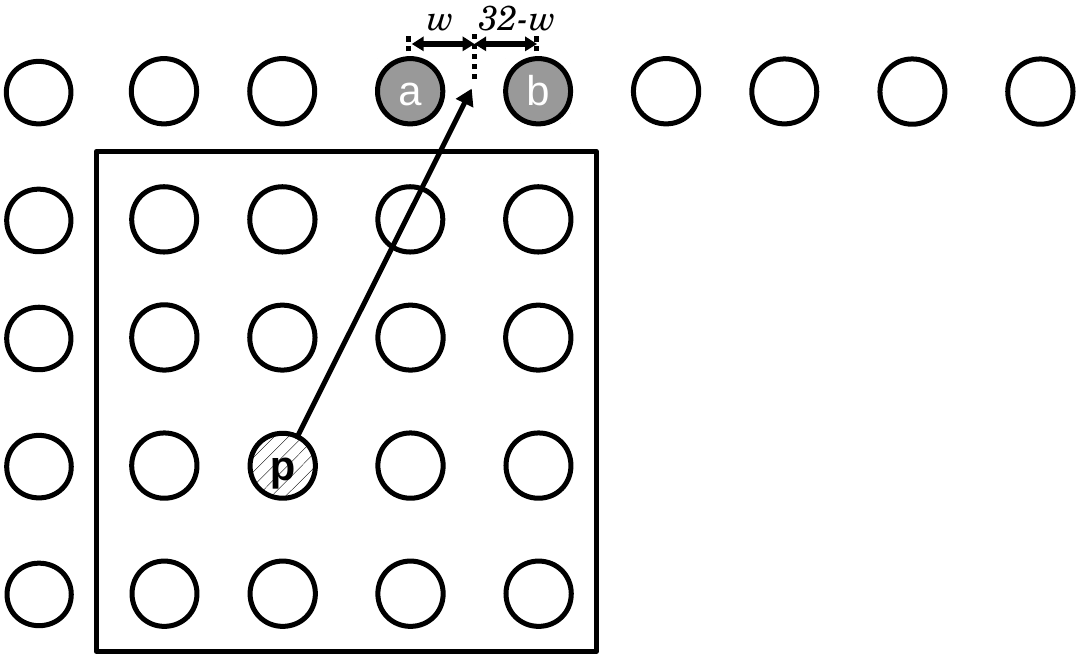}}
  \caption{Block-based angular prediction in HEVC. Block pixel to be predicted is projected on the block neighbor pixels (reference samples) along an angular direction. Prediction $p$ is obtained from linear interpolation of two closest reference samples $a$ and $b$.}
\label{fig:ang_interp}
\end{figure}

\begin{figure}[tb]
  \centering
  \centerline{\includegraphics[scale=0.48, trim = 1.2cm 10.0cm 1.1cm 1.2cm, clip=true]{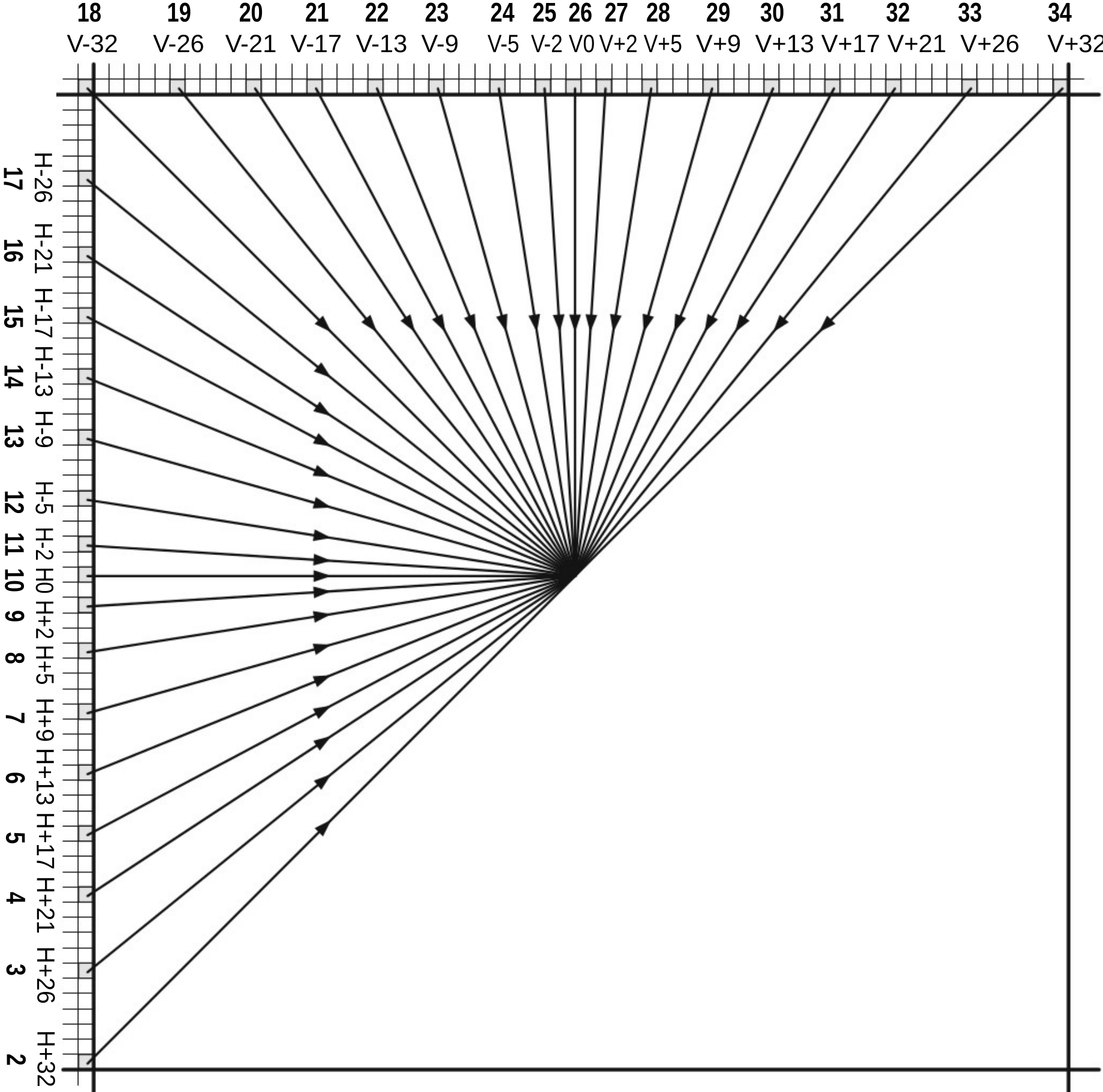}}
  \caption {HEVC angular intra prediction modes numbered from 2 to 34 and the associated displacement parameters. H and V indicate the horizontal and vertical directionality, respectively, and the numeric part of the parameter refers to the pixels’ displacement as 1/32 pixel fractions. (picture reproduced from \cite{IntraHEVC})}
\label{fig:intra_modes}
\end{figure}

In the angular modes, the location of a block pixel to be predicted is projected to the reference samples (i.e. neighbor pixels of the block) along an angle, as shown in Figure \ref{fig:ang_interp}. Each of the 33 angular prediction modes uses a different projection angle, as shown in Figure \ref{fig:intra_modes}. The two closest pixels to the projected location in the reference samples are used to linearly interpolate (at 1/32 pixel accuracy) a prediction value as follows :
\begin{equation}
\label{eq:ang_pre}
p=((32-w) \cdot a + w \cdot b+16)>>5.
\end{equation}
Here "$>>$" indicates a bit shift operator, $a$ and $b$ represent the reference samples, and $32-w$ and $w$ represent 5-bit integer interpolation weights, which are determined by the angle or prediction mode \cite{IntraHEVC} (See Figure \ref{fig:ang_interp}). 

\subsection{Spatial prediction methods based on residual differential pulse code modulation (RDPCM)}
\label{ssec:rdpcm}
Since the transform is skipped in losless intra coding, the block-based spatial prediction alone can not efficiently reduce the spatial correlation. Therefore, spatial prediction methods based on residual differential pulse code modulation (RDPCM) first perform the default block-based intra prediction and then process the prediction error block further with a second pixel-by-pixel prediction step. 

There are many methods proposed in the literature based on this approach. One of the earliest of such methods was proposed in \cite{sulivanDPCM} for lossless intra coding in H.264/AVC. Here, first the block-based spatial prediction is performed, and then a simple pixel-by-pixel differencing operation is applied on the residual pixels in only horizontal and vertical intra prediction modes. In horizontal mode, from each residual pixel, its left neighbor is subtracted and the result is the RDPCM pixel of the block. Similar differencing is performed along the vertical direction in the vertical intra mode. Note that the residuals of other angular modes are not processed in \cite{sulivanDPCM}. This method was later accepted into H.264/AVC extension profile due to its simplicity, good interplay with block-based prediction and its good performance \cite{sullivan2004h}. Note also that this method is sometimes referred to as the RDPCM method in the lossless intra coding literature although it is one possible RDPCM approach.

Another method based on RDPCM is proposed for HEVC in \cite{cross} and is termed cross RDPCM (CRDPCM). The first part of this method is the same as in the above RDPCM method. In other words, first block-based spatial prediction is performed, and then for only horizontal and vertical modes, the prediction residual block is further processed with vertical pixel-by-pixel differencing in the vertical mode and horizontal pixel-by-pixel differencing in the horizontal mode. In the second part of CRDPCM method, another pixel-by-pixel differencing is applied on the obtained RDPCM pixels along the cross direction. In other words, in the horizontal intra mode, the RDPCM pixels are subtracted from their upper neighbors along the vertical direction, and in the vertical mode, the RDPCM pixels are subtracted from their left neighbors along the horizontal direction. Note however that this second part is not always applied, but only in the blocks where it provides further compression, which is determined at the encoder with rate-distortion (RD) optimized decision and signaled to the decoder with a flag in the bitstream. 

Another proposed method similar to CRDPCM is secondary RDPCM (SRDPCM) \cite{Secondary}. This method applies the second and optional differencing operation on the RDPCM pixels not along the cross direction, but along the same direction. Another method similar to CRDPCM and SRDPCM is proposed in \cite{IRDPCM}. This method applies the second and optional differencing operation on the RDPCM pixels not along the cross or same direction but along both directions by subtracting from the RDPCM pixel the average of its left and upper neighbors. Yet another similar method is proposed in \cite{ARDPCM}. Here, instead of processing the block-based prediction residual with a pixel-by-pixel differencing along horizontal or vertical directions, a general linear prediction is applied using three neighbor residual pixels. The method is applied to all intra modes of HEVC and the linear prediction weights are updated at the encoder/decoder during encoding/decoding from previously encoded/decoded pixels.


While the simple RDPCM method \cite{sulivanDPCM} provides modest improvements in coding gain with respect to only block-based prediction, more complicated methods such as CRDPCM \cite{cross} can require additional RD-optimized decisions at the encoder and syntax change to signal decision. 

\subsection{Spatial prediction methods based on pixel-by-pixel prediction}
\label{ssec:sap}
When the transform step is skipped in lossless intra coding, the block-based spatial prediction becomes less effective since some block pixels are predicted from distant reference samples and there is no transform step that can compensate for this inefficient prediction. However, since the transform is skipped in lossless coding, a pixel-by-pixel spatial prediction approach can now be used instead of a block-based approach for more efficient prediction. 

\begin{figure}[tb]
  \centering
  \centerline{\includegraphics[scale=0.65, trim = 5.1cm 21.0cm 4.9cm 2.0cm, clip=true]{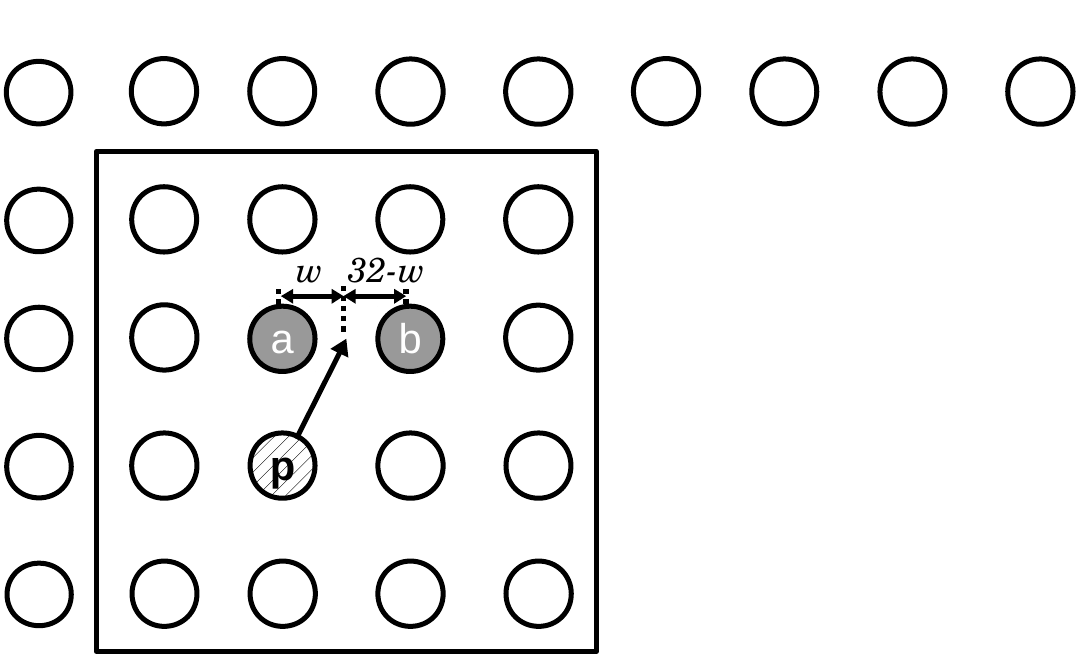}}
  \caption{Sample-based Angular Prediction (SAP) proposed for HEVC in \cite{SAP}. Block pixel to be predicted is projected on the immediately above row (or column, depending on mode) pixels (reference samples) along an angular direction. Prediction $p$ is obtained from linear interpolation of two closest reference samples $a$ and $b$.}
\label{fig:sap}
\end{figure}

Many lossless intra coding methods based on pixel-by-pixel prediction  approach appeared in the literature \cite{SAP,DC2,Templatebased,ADSAP}. One such method applied in HEVC is provided in \cite{SAP} and is called Sample-based Angular Prediction (SAP). In the SAP method, the planar and DC modes are not modified. In the angular modes, the same angular projection directions and linear interpolation equations of HEVC are used, and only the used reference samples are modified. Instead of the the block neighbor pixels, the immediate neighbor pixels are used as reference pixels, as shown in Figure \ref{fig:sap}, resulting in a pixel-by-pixel prediction version of the HEVC block-based spatial prediction approach. 

Another pixel-by-pixel prediction approach is given in \cite{DC2}, where it is applied to only the DC prediction mode. In this method, called piecewise DC prediction (PWDC), each block pixel is predicted by the average of its left and upper neighbors in the DC mode. 
The authors of \cite{DC2} also combine their approach with the RDPCM and CRDPCM methods discussed in Section \ref{ssec:rdpcm} and report improved coding gains.

Another pixel-by-pixel prediction approach is given in \cite{Templatebased}, where it is applied to only the planar prediction mode. Each block pixel is predicted by a weighted average of its four neighbors (left, upper-left, upper, upper-right) and the weights are determined during encoding/decoding using a table look-up \cite{Templatebased}. 

A more recent pixel-by-pixel prediction approach is given in \cite{ADSAP}, and is called adaptive directional SAP (AD-SAP). This method is similar to the SAP method discussed above, in the sense that the same prediction equations are used in a pixel-by-pixel manner. However, unlike the SAP method which uses the same prediction direction inside a single prediction block, the AD-SAP method may adaptively change the prediction direction for each pixel inside a prediction block, by checking how well the neighbor pixels can be predicted from their respective neighbor pixels \cite{ADSAP}.

While the pixel-by-pixel prediction methods discussed here provide improved coding gains, \cite{SAP} applies the method to only angular modes, \cite{DC2} applies it to only DC mode, \cite{Templatebased} applies it to only planar mode with a complex algorithm and \cite{ADSAP} requires complex calculations to change prediction direction for each pixel. As will be discussed in Section \ref{sec:pro_app}, this paper provides a unified treatment of pixel-by-pixel prediction applied to all modes with optimized prediction weights and can provide better coding gains.

\subsection{Methods based on modified entropy coding}
\label{ssec:mod_ent}
In lossy coding, transform coefficients of prediction residuals are entropy coded, while in lossless coding, the prediction residuals are entropy coded. Considering the difference of the statistics of quantized transform coefficients and prediction residuals, several modifications in entropy coding were proposed for lossless coding in HEVC \cite{kim2010efficient,choi2013differential,gao2011lossless,CABAC3,piao2013ahg7,CABAC1}. A brief overview is as follows. In \cite{choi2013differential}, the scanning order is reversed and the binarization used in the entropy coder is modified. An intra mode dependent scanning order is proposed in \cite{gao2011lossless}. Coding of the last position flag is modified in \cite{CABAC3,piao2013ahg7} and coding of the levels of the coefficients/residuals is modified in \cite{CABAC1}.

\section{Lossless Intra Coding with 3-tap Filters}
\label{sec:pro_app}
This section presents the spatial prediction method explored in this paper for lossless intra coding in HEVC. The method is based on the pixel-by-pixel prediction approach discussed in Section \ref{ssec:sap}.

\subsection{3-tap filtering approach}
\label{ssec:3tap}
A simple way to obtain a pixel-by-pixel prediction method for lossless intra coding in HEVC is to slightly modify its block-based angular  prediction method so that the reference samples used for prediction are not taken from the distant block neighbor pixels but from the immediate neighbors in the above row or left column, depending on mode. This is exactly what is done in the SAP method proposed in \cite{SAP}, which was  discussed in Section \ref{ssec:sap}.

\begin{figure}[tb]
  \centering
  \centerline{\includegraphics[scale=0.65, trim = 5.1cm 21.0cm 4.9cm 2.0cm, clip=true]{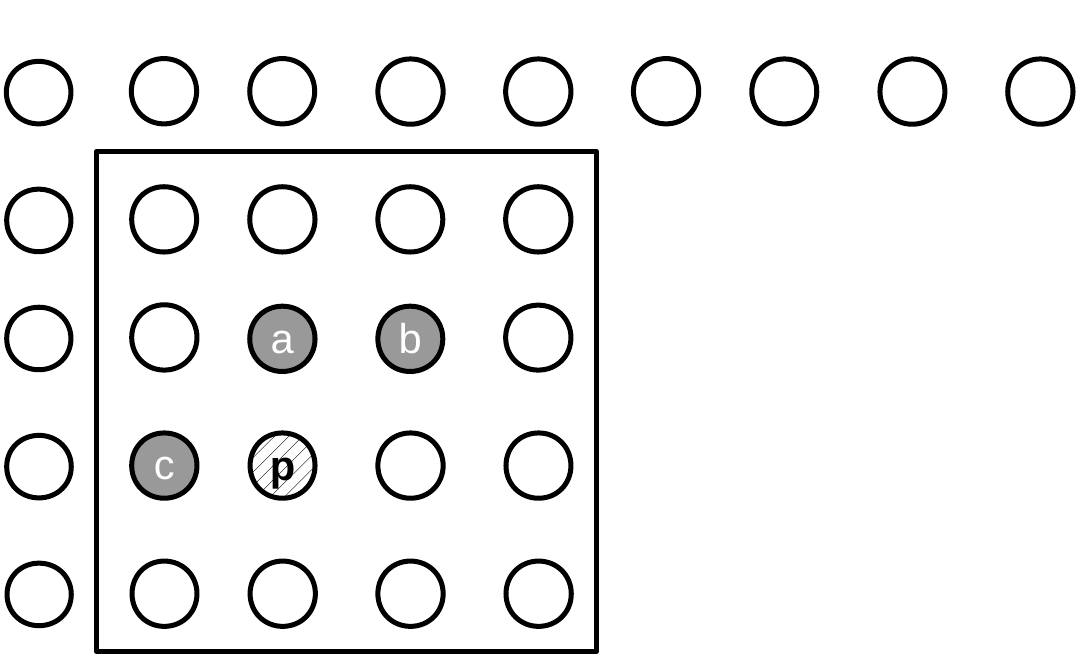}}
  \caption{Proposed 3-tap filtering for lossless intra coding in HEVC. Each block pixel is predicted from a two-dimensional neighborhood of pixels based on a two-dimensional correlation model. Prediction $p$ is obtained as a linear combination of neighbor pixels $a$, $b$ and $c$.}
\label{fig:3tap}
\end{figure}

A major reason why the angular projection based spatial prediction method is commonly used for block-based spatial prediction is that it is a computationally simple method and works well together with a transform and adaptive block-sizes in lossy intra coding. However, the angular projection approach assumes a strong directional and one-dimensional correlation among pixels, which is a limited model. In particular, in a pixel-by-pixel neighborhood used in lossless intra coding, a two-dimensional pixel correlation (although possibly directional) can be a much better model and does not increase computational complexity of prediction significantly. As a result, this paper proposes a pixel-by-pixel prediction approach where each block pixel is predicted from a two-dimensional neighborhood of pixels. An example is shown in Figure \ref{fig:3tap} where three neighbor pixels (left, upper, upper-right) are used for predicting a block pixel. The prediction in Figure \ref{fig:3tap} can be compared to the prediction of \cite{SAP} shown in Figure \ref{fig:sap}. Note that similar two-dimensional correlation models were also used for lossy intra coding with improved results over block-based angular projection methods \cite{MrkvIntra,Rose1}.

\begin{figure}[tb]
\begin{minipage}[b]{0.21\linewidth}
  \centering
  \centerline{\includegraphics[scale=0.70, trim = 6.0cm 19.5cm 13.2cm 4.7cm, clip=true]{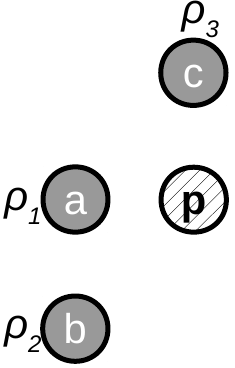}}
  \begin{footnotesize}\centerline{(a) Modes 2-9}\end{footnotesize}\medskip
\end{minipage}
\begin{minipage}[b]{0.25\linewidth}
  \centering
  \centerline{\includegraphics[scale=0.70, trim = 6.0cm 19.5cm 13.2cm 4.7cm, clip=true]{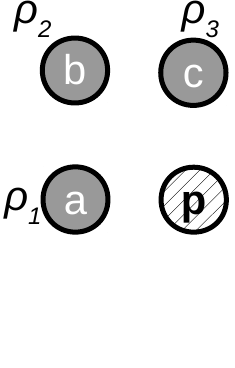}}
  \begin{footnotesize}\centerline{(b) Modes 0,1,10-18}\end{footnotesize}\medskip
\end{minipage}
\begin{minipage}[b]{0.25\linewidth}
  \centering
  \centerline{\includegraphics[scale=0.70, trim = 6.0cm 19.5cm 13.2cm 4.7cm, clip=true]{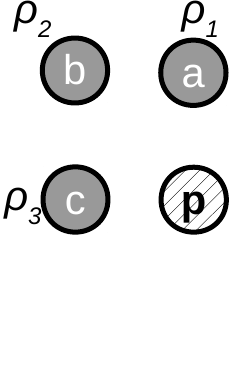}}
  \begin{footnotesize}\centerline{(c) Modes 19-26}\end{footnotesize}\medskip
\end{minipage}
\begin{minipage}[b]{0.26\linewidth}
  \centering
  \centerline{\includegraphics[scale=0.70, trim = 6.0cm 19.5cm 11.9cm 4.7cm, clip=true]{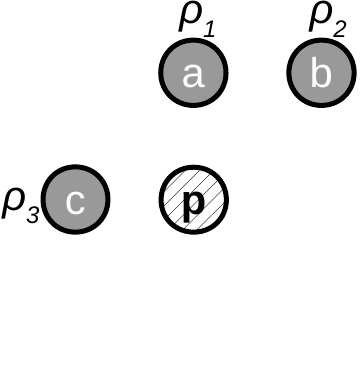}}
  \begin{footnotesize}\centerline{(d) Modes 27-34}\end{footnotesize}\medskip
\end{minipage}
\caption{Neighbor pixels used for prediction in the 3-tap filtering method according to intra modes of HEVC. Intra modes 2-9 use neighbor pixels shown in (a), planar, DC modes (0,1) as well as intra modes 10-18 use neighbor pixels shown in (b), intra modes 19-26 use neighbor pixels shown in (c) and intra modes 27-34 use neighbor pixels shown in (d).}
\label{fig:3taps}
\end{figure}

The pixels in the two-dimensional neighborhood to be used for prediction must be chosen carefully to reflect various possible directional correlation in images. In this paper, the neighbor pixels shown in Figure \ref{fig:3taps} are used for prediction in different intra modes of HEVC. These neighbor pixels were chosen so that each intra mode in 3-tap filtering method can reflect similar directionality as the same mode in HEVC. The prediction equation in each mode is the same and is given in Equation (\ref{eq:3taps}) below, however, the location of the neighbor pixels $a$, $b$ and $c$ as well as the prediction weights $\rho_1$, $\rho_2$ and $\rho_3$ change according to intra mode as shown in Figure \ref{fig:3taps}.

\begin{equation}
\label{eq:3taps}
p = \rho_1 \cdot a + \rho_2 \cdot b + \rho_3 \cdot c.
\end{equation}

Notice that unlike in the SAP method \cite{SAP}, the proposed pixel-by-pixel prediction approach is used for all angular modes as well as the DC and planar modes. Notice also that it is possible to use four or even five neighbors for prediction but due to the large number of prediction modes in HEVC, the increasing computations and diminishing coding gains, only three neighbors are used for prediction in all intra modes. 

\subsection{Prediction weights}
\label{ssec:pred_w}
\newcommand{\argmin}{\operatornamewithlimits{argmin}}
\newcommand{\argmax}{\operatornamewithlimits{argmax}}
While Equation (\ref{eq:3taps}) provides the prediction expression, a major question is how to find the weights $\rho_1$, $\rho_2$ and $\rho_3$ for each intra mode of lossless intra coding in HEVC. These weights were determined from a training sequence\footnote{A training sequence was formed from several images in the JPEG-XR image test set \cite{JPEGXR}.} using a two-stage optimization approach. In the first stage, an iterative minimum-squared-error (MSE) method is used, which finds the weights that minimize the mean squared prediction error over the training sequence. These parameters are then further refined in the second stage by minimizing the bitrate over the training sequence. These final weights will then be later used to code the test sequences of HEVC.

In the first stage of the optimization approach, the weights $\rho_1$, $\rho_2$ and $\rho_3$ are obtained offline from a training sequence by minimizing the squared-error, i.e. the sum of squares of prediction errors in the training sequence. A prediction error pixel $r$ in a block is obtained by subtracting out, from the original block pixel $o$, the prediction $p$ given in Equation (\ref{eq:3taps}) :
\begin{align}
r = & o - p \nonumber \\ 
  = & o - ( \rho_1 \cdot a + \rho_2 \cdot b + \rho_3 \cdot c ).
\label{eq:mse_err}
\end{align}
The weights $\rho_1$, $\rho_2$ and $\rho_3$ are then found by minimizing the sum of the squares of prediction error $r$ for all pixels from the training sequence that were coded with the same intra mode :
\begin{align}
\argmin_{\rho_1,~ \rho_2,~ \rho_3}  \sum_{I_k} ( o - \rho_1 \cdot a - \rho_2 \cdot b - \rho_3 \cdot c ) ^{2} .
\label{eq:mse_min} 
\end{align}
Note that the above sum is over the set $I_k$ which includes all pixels in the training sequence coded with the same intra mode. In other words, the above minimization problem is solved for each intra mode separately. The solution can be easily obtained and is given by Equation (\ref{eq:mse_mat}):
\begin{align}
\begin{bmatrix} \hat{\rho_1} \\ \hat{\rho_2} \\ \hat{\rho_3}  \end{bmatrix} 
=
\begin{bmatrix} 
\sum\limits_{I_k} a^2         & \sum\limits_{I_k} a\cdot b  & \sum\limits_{I_k} a\cdot c  \\[.4cm]
\sum\limits_{I_k} b\cdot a    & \sum\limits_{I_k} b^2       & \sum\limits_{I_k} b\cdot c  \\[.4cm]
\sum\limits_{I_k} c\cdot a    & \sum\limits_{I_k} c\cdot b  & \sum\limits_{I_k} c^2  \\[.4cm] 
\end{bmatrix}
^{-1} 
\begin{bmatrix} \sum\limits_{I_k} o\cdot a \\[.4cm] \sum\limits_{I_k} o\cdot b \\[.4cm] \sum\limits_{I_k} o\cdot c \end{bmatrix}
\label{eq:mse_mat}
\end{align}

Finally, to obtain the weights for all intra modes collectively, an iterative approach is used. In each iteration, the training sequence are encoded using the weights from the previous iteration, and Equation (\ref{eq:mse_mat}) is solved at the end for each intra mode to obtain the new weights for the next iteration. The initial iterations uses the SAP method \cite{SAP}, and the iterations are stopped when the estimated weights don't change significantly and the coding gain stabilizes, which typically takes about five to ten iterations. 

In the second stage of the optimization approach, the prediction weights from the first stage are further refined by minimizing the bitrate over the training sequence. While MSE and bitrate are in general coherent, the ultimate performance metric of a lossless codec is the bitrate, and therefore a second-stage optimization is performed to further fine-tune the weights to achieve minimum bitrate instead of MSE over the training sequence. The exact optimization method used here will be discussed in Section \ref{ssec:optbit}.

\subsection{Implementation aspects}
\label{ssec:imp}
While Sections \ref{ssec:3tap} and \ref{ssec:pred_w} discuss the proposed pixel-by-pixel prediction approach and the determination of the prediction weights, a number of other aspects related to the implementation and simplification of the proposed prediction method need to be carefully considered. These aspects are the bit-depth of the weights $\rho_1$, $\rho_2$ and $\rho_3$, whether the weights in each intra mode should be changed depending on the block-sizes available in HEVC, and whether the same weights can be used for symmetric modes such as horizontal and vertical modes. Note that the analysis presented here uses only the MSE method to determine the prediction weights. The second-stage optimization method will be used in Section \ref{ssec:optbit} based upon the results provided here.

\subsubsection{Weights in different block-sizes}
While the default block-based spatial prediction in HEVC and the pixel-by-pixel SAP  method in \cite{SAP} use the same interpolation weights at all block-sizes in each intra mode, it needs to be examined how the compression performance changes in the proposed prediction method with different weights in each block-size. To examine this, mode-dependent weights are obtained for each PU block-size available in HEVC (4x4 to 32x32) separately and together. To obtain separate weights in each block-size, the set of training sequence pixels $I_k$ in Equations (\ref{eq:mse_err}) and (\ref{eq:mse_min}) is formed separately for each intra mode and each block-size, resulting in a total of $35\times 4=140$ different mode and block-size dependent weights. To use same weights in each block-size, the set of training sequence pixels $I_k$ in Equations (\ref{eq:mse_err}) and (\ref{eq:mse_min}) is formed for each intra mode using all available block-sizes, resulting in a total of $35$ different and only mode-dependent weights. The experimental results provided in Section \ref{sec:exp_res} show that only mode dependent weights are sufficient and allowing the weights to change across the block-size does not bring significant coding gains.

\subsubsection{Bit-depth of weights}
While the default block-based spatial prediction in HEVC and the pixel-by-pixel SAP  method in \cite{SAP} use 5-bits for the prediction weights (in interpolation at 1/32 pixel  fractions), it needs to be examined how varying the bit-depth of the weights affects the compression performance in the proposed prediction method. To examine this, the weights obtained as discussed in Section \ref{ssec:pred_w} are quantized to 10-bits and 5-bits, and obtainable coding gains with both bit-depths are compared in the experimental results in Section \ref{sec:exp_res}. These results show that using 5-bits for the weights is sufficient and the coding gain drop from using 10-bit weights is insignificant.

\subsubsection{Weights in symmetric modes}
While the default block-based spatial prediction in HEVC and the pixel-by-pixel SAP  method in \cite{SAP} use the same interpolation weights for symmetric modes such as modes 17 and 19 (see Figure \ref{fig:intra_modes}), it needs to be examined how using different or same weights in the symmetric modes affects the compression performance in the proposed prediction method. To examine this, 5-bit and only mode-dependent weights are obtained separately and together for each symmetric mode. To obtain separate weights in each mode, the set of training sequence pixels $I_k$ in Equations (\ref{eq:mse_err}) and (\ref{eq:mse_min}) is formed separately for each intra mode, resulting in a total of $35$ different mode dependent weights. To use same weights in symmetric modes, the set of  training sequence pixels $I_k$ in Equations (\ref{eq:mse_err}) and (\ref{eq:mse_min}) is formed for each pair of symmetric intra modes using pixels coded from both modes, resulting in a total of $1+1+1+16=19$ different weights. (Note that DC, planar and mode 18 do not have symmetric pairs and only the remaining 16 modes have symmetric pairs.) The experimental results provided in Section \ref{sec:exp_res} show that symmetric  weights are sufficient and allowing different weights for symmetric modes does not bring significant coding gains.

\section{Experimental Results}
\label{sec:exp_res}
This section provides experimental results to compare the proposed pixel-by-pixel prediction approach based on 3-tap filtering against other approaches, in terms of achievable compression efficiency and incurred computational complexity. First, Section \ref{ssec:exp_set} presents the experimental settings used in all provided experimental results. Next, Section \ref{ssec:analysis} presents experimental results to analyze the implementation aspects of the proposed 3-tap filtering method discussed in Section \ref{ssec:imp}. Then, Section \ref{ssec:optbit} discusses fine-tuning the 3-tap filter weights by optimizing for bitrate and presents improved results. Next, Section \ref{ssec:compar} compares the compression efficiency and some statistical analysis results of the 3-tap filtering method with default HEVC and SAP methods. Finally, Section \ref{ssec:times} provides a complexity analysis of the compared methods.

\subsection{Experimental settings}
\label{ssec:exp_set}
The proposed pixel-by-pixel prediction approach based on 3-tap filtering is implemented into the HEVC reference software HM12.0 \cite{HM12}. For the provided experimental results, the common test conditions and the AI-Main reference configuration (in which all frames are coded as intra frames) in \cite{commontest} are followed. All test sequences in class A to F are used in the experiments.

Notice that in all the experimental results presented below, the used prediction weights were obtained offline as discussed in Section \ref{ssec:pred_w} from a training sequence that does not include any of the experiment sequences.

\subsection{Analysis of implementation aspects of the proposed 3-tap filtering method}
\label{ssec:analysis}
This section presents experimental results to analyze the implementation aspects of the proposed 3-tap filtering method discussed in Section \ref{ssec:imp}. In particular, the analyzed aspects are the bit-depth of the prediction weights, whether the weights in each intra mode should be changed depending on the block-sizes available in HEVC, and whether the same weights can be used for symmetric modes.

First, prediction weights are obtained in the most general case from the training sequences. In this case, the weights were allowed to change in different PU block-sizes, have a large bit-depth (10-bits), and were not required to be same in symmetric modes. In this most general case, a total of $35\times 4=140$ different mode and block-size dependent 10-bit weights were obtained from the training sequence and used in the experiments. The compression results are shown in the first column of Table \ref{tb:3tap} as percentage bit-rate reduction over the default block-based lossless intra coding in HEVC. It can be seen that significant bitrate savings can be achieved with an average of $10.84\%$.

Next, to analyze how varying the weights in different block-sizes affects the compression performance in the proposed prediction method, the weights at different block-sizes were forced to be same in each intra mode, resulting in a total of $35$ only mode-dependent 10-bit weights. The compression results of this case are shown in the second column of Table \ref{tb:3tap}. It can be seen that the compression results hardly change. In fact, the average bitrate reduction goes up by $0.02\%$ to $10.86\%$.

Next, to analyze the effect of the bit-depth of the prediction weights, the obtained $35$ mode-dependent weights were quantized to 5-bits to be congruent with HEVC interpolation weights. The compression results of this case are shown in the third column of Table \ref{tb:3tap}. It can be seen that the average bitrate reduction drops slightly to $10.70\%$.

Finally, to analyze the effect of using same weights in symmetric modes, the weights for symmetric modes were forced to be the same, resulting in a total of $19$ mode-dependent 5-bit weights. The compression results of this final case are shown in the final column of Table \ref{tb:3tap}. It can be seen that the average bitrate reduction drops slightly to $10.54\%$.

To summarize, the presented experimental results indicate that the discussed implementation aspects do not affect the compression performance in a significantly unfavorable way, and the simplest and most implementation friendly case, where only 19 mode-dependent 5-bit weights are used, achieves a competitive $10.54\%$ bitrate reduction over block-based HEVC lossless intra coding.

\begin{table}[h]
\centering
\caption{Average percentage bitrate reduction of proposed 3-tap filtering method with different implementation settings over HEVC lossless intra coding.}
\label{tb:3tap}
\begin{tabular}{l|c|c|c|c}
\hline
  & General      & Block-size                    & + Bit-depth                   & + Symmetry          \\
  &     case     &            dependency         &             reduced to        &            enforced \\ 
  &              &                       removed &                        5-bits &                     \\ \hline \hline
Class A   & 15.52      & 15.42      & 15.30     & 15.28          \\ \hline
Class B   & 7.71       & 7.83       & 7.63      & 7.21           \\ \hline
Class C   & 8.62       & 8.61       & 8.48      & 8.40           \\ \hline
Class D   & 10.65      & 10.72      & 10.58     & 10.47          \\ \hline
Class E   & 13.02      & 13.00      & 12.82     & 12.70          \\ \hline
Average   & 10.84      & 10.86      & 10.70     & 10.54          \\ \hline
\end{tabular}
\end{table}

\subsection{Fine-tuning of weights of 3-tap filtering method by optimizing for bitrate}
\label{ssec:optbit}
While prediction MSE and coding bitrate are in general coherent, in the sense that smaller squared prediction error generally leads to smaller coding bitrate, obtaining prediction weights by minimizing squared prediction error is mismatched with the ultimate bitrate that lossless coders optimize. In other words, the prediction weights which minimize squared prediction error may not necessarily produce minimum bitrate.  Hence, a second-stage optimization is performed to further fine-tune the weights from the MSE optimization stage to achieve minimum bitrate instead of MSE over the training sequence.

Based on the results presented in the previous section (Section \ref{ssec:analysis}), only the simplest and most implementation friendly case with only 19 mode-dependent 5-bit weights is further refined by optimizing for bitrate. A simple optimization algorithm is used to fine-tune the prediction weights. 

First, the HEVC coder is run with the prediction weights obtained from first stage (MSE) optimization and the resulting bitrate is $B_{opt}$. Let $\rho_{1,k}$, $\rho_{2,k}$ and $\rho_{3,k}$ be the prediction weights of mode $k$ and its symmetric mode if it exists. Then, the optimization approach consists of applying the following steps : 
\begin{enumerate}
  \item Let $k=0$ and $B_{best}=B_{opt}$.
  \item Generate 6 candidates for prediction weights of mode $k$ ($\rho_{1,k,i}$ , $\rho_{2,k,i}$ and $\rho_{3,k,i}$), run HEVC coder by replacing mode $k$'s weights with the candidates and record the resulting bitrates $B_i$.
\begin{table}[h!]
  \caption{Candidate prediction weights}
  \label{tb:tesseq}
  \centering
\begin{tabular}{c|c|c|c|c} 
  \hline   
 (Candidate) i & $\rho_{1,k,i}$   & $\rho_{2,k,i}$   & $\rho_{3,k,i}$   & Bitrate $B_i$ \\ 
  \hline \\[-2.0ex] \hline
             1 & $\rho_{1,k}+1$ & $\rho_{2,k}-1$ & $\rho_{3,k}$   & $B_1$  \\  \hline
             2 & $\rho_{1,k}-1$ & $\rho_{2,k}+1$ & $\rho_{3,k}$   & $B_2$  \\  \hline
             3 & $\rho_{1,k}+1$ & $\rho_{2,k}$   & $\rho_{3,k}-1$ & $B_3$  \\  \hline
             4 & $\rho_{1,k}-1$ & $\rho_{2,k}$   & $\rho_{3,k}+1$ & $B_4$  \\  \hline
             5 & $\rho_{1,k}$   & $\rho_{2,k}+1$ & $\rho_{3,k}-1$ & $B_5$  \\  \hline
             6 & $\rho_{1,k}$   & $\rho_{2,k}-1$ & $\rho_{3,k}+1$ & $B_6$  \\  \hline 
\end{tabular}
\end{table}
Find candidate with smallest bitrate, $i^{*}=\argmin_{i} B_i$. If $B_{i^*}<B_{opt}$, update bitrate $B_{opt}=B_{i^*}$ and prediction weights for mode $k$ and its symmetric mode $\rho_{1,k}=\rho_{1,k,i^*}$ , $\rho_{2,k}=\rho_{2,k,i^*}$ and $\rho_{3,k}=\rho_{3,k,i^*}$.
\item If $k<18$, (i.e. not the last mode without a symmetric pair), increment $k$ by one and go to step 2. If $k=18$, check if this iteration over all intra modes improved bitrate, i.e. $B_{opt}<B_{best}$. If so, go to step 1, otherwise finish.
\end{enumerate}

The prediction weights obtained from this algorithm increase the bitrate reduction over HEVC further as shown in Table \ref{tb:opt}. Table \ref{tb:w_3tap} lists the 19 mode-dependent 5-bit weights of the final obtained 3-tap filters.

\begin{table}[tbh]
\centering
\caption{Average percentage bitrate reduction of the proposed 3-tap filtering method over HEVC lossless intra coding after first (for MSE) and second stage (for bitrate) of optimizations.}
\label{tb:opt}
\begin{tabular}{l|c|c}
\hline
          & After optimization & After optimization                 \\ 
          & for MSE            & for bitrate \\ \hline \hline
Class A   & 15.28     &  15.67   \\ \hline
Class B   & 7.21      &   7.81  \\ \hline
Class C   & 8.40      &   8.88  \\ \hline
Class D   & 10.47     &  11.10   \\ \hline
Class E   & 12.70     &  14.38   \\ \hline
Average   & 10.54     &  11.24   \\ \hline
\end{tabular}
\end{table}

\begin{table}[tbh]
\centering
      \caption{3-tap filter weights}
      \centering
      \label{tb:w_3tap}
\begin{tabular}{c|c|c|c}
\hline
Modes & $\rho_1$ & $\rho_2$ & $\rho_3$    \\ \hline \hline
0     & 22  & -11 & 21 \\ \hline
1     & 19  & -1  & 14 \\ \hline
2,34  & -11 & 29  & 14 \\ \hline
3,33  &  0  & 22  & 10 \\ \hline
4,32  & 10  & 22  & 0  \\ \hline
5,31  & 10  & 14  & 8  \\ \hline
6,30  & 25  & 12  & -5 \\ \hline
7,29  & 19  &  4  & 9  \\ \hline
8,28  & 29  & 5   & -2  \\ \hline
9,27  & 31  & -2  & 3  \\ \hline
10,26 & 30  & -25 & 27 \\ \hline
11,25 & 32  & -11 & 11 \\ \hline
12,24 & 27  & -16 & 21 \\ \hline
13,23 & 23  & 0   & 9  \\ \hline
14,24 & 15  & 6   & 11  \\ \hline
15,21 & 22  & 14  & -4  \\ \hline
16,20 & 14  & 22  & -4  \\ \hline
17,19 & 5   & 29  & -2 \\ \hline
18    & 7   & 14  & 11  \\ \hline
\end{tabular}
\end{table}

\subsection{Comparison to HEVC, SAP and other methods}
\label{ssec:compar}
The compression efficiency of the proposed 3-tap filtering method is compared against HEVC, the SAP method \cite{SAP}, and several other methods from the literature. Table \ref{tb:compar} shows the results. It can be seen from the table that the proposed 3-tap filtering method achieves the best average coding gain result; an average $11.34\%$ bitrate reduction over HEVC. Notice that we implemented only the SAP method in addition to our proposed 3-tap filtering method, while the results of the other methods are taken from the respective papers.


\newcolumntype{P}[1]{>{\centering\arraybackslash}p{#1}}

\begin{table*}[tbh]
\centering
\caption{Average percentage bitrate reduction of multiple methods from the literature and the proposed 3-tap filtering method over HEVC lossless intra coding.}
\label{tb:compar}
\begin{tabular}{l|P{2cm}|P{2cm}|P{2cm}|P{2cm}|P{2cm}|P{2.0cm}}
\hline 
                  & RDPCM & CRDPCM & PWDC  & SAP   & AD-SAP   & Proposed \\
                  &\cite{cross}&\cite{cross} &\cite{DC2}&\cite{SAP}&\cite{ADSAP}   & 3-tap filtering \\ \hline \hline
Class A           & 7.19  & 11.3   & 10.06 & 8.7   & 11.11   & 15.67 \\ \hline
Class B           & 3.54  & 3.91   & 5.29  & 5.11  & 5.45    & 7.81  \\ \hline
Class C           & 4.46  & 4.76   & 4.29  & 6.99  & 6.47    & 8.88  \\ \hline
Class D           & 6.3   & 6.91   & 4.89  & 8.65  & 8.85    & 11.1  \\ \hline
Class E           & 8.32  & 9.72   & 6.74  & 10.54 & 10.84   & 14.38 \\ \hline
Class F           & 9.82  & 10.11  & 4.77  & 12.44 & 13.75   & 11.81 \\ \hline
Average           & 6.12  & 8.43   & 5.95  & 8.51  & 9.41    & 11.34 \\ \hline
\end{tabular}
\end{table*}

\begin{figure*}[tbh] 
\centering 
\begin{tikzpicture} 
\begin{axis}
            [
                compat=newest, ymajorgrids, ybar, width=1.00\textwidth, height=5.0cm, ymin=0, ymax=20, bar width=0.05cm, xmin=-.5, xmax=34.5,
                ylabel = {Frequency (\%)}, xlabel = {Intra mode},
                legend entries={HEVC , SAP \cite{SAP}, 3-tap filter},
	              legend image code/.code={	\draw[#1] (-0cm,-0cm) rectangle (0.4cm,0.15cm);},
	              legend style={at={(0.80,+0.55)},anchor=south west}
            ] 
\addplot
	coordinates {(0,12.49)(1,8.25)(2,1.79)(3,1.37)(4,1.60)(5,2.09)(6,	2.70)(7,	2.96)(8,	2.90)(9,	2.91)(10,	7.29)(11	,3.23)(12,	2.52)(13,	2.41)(14,	1.94)(15,	1.79)(16,	1.69)(17,	1.79)(18,	1.47)(19,	1.81)(20,	1.64)(21,	1.71)(22,	1.80)(23,	2.42)(24,	2.37)(25,	3.13)(26,	7.54)(27,	1.91)(28,	2.17)(29,	2.11)(30,	1.81)(31,	1.57)(32,	1.37)(33,	1.46)(34,	1.99)};
\addplot 
	coordinates {(0,3.33)(1,3.02)(2,0.43)(3,0.66)(4,1.25)(5,2.21)(6,	4.57)(7,6.24)(8,	5.66)(9,	3.87)(10,	3.08)(11	,3.43)(12,	5.13)(13,	5.30)(14,	3.55)(15,	1.91)(16,	1.08)(17,	0.66)(18,	0.26)(19,	0.72)(20,	1.22)(21,	1.84)(22,	3.30)(23,	5.05)(24,	5.50)(25,	3.84)(26,	3.99)(27,	3.02)(28,	5.01)(29,	4.79)(30,	2.77)(31,	1.63)(32,	0.82)(33,	0.52)(34,	0.30)};
\addplot 
	coordinates {(0,8.49)(1,5.39)(2,0.89)(3,1.15)(4,1.28)(5,1.97)(6,	2.46)(7,3.04)(8,	2.26)(9,	3.34)(10,	16.44)(11	,3.39)(12,	1.99)(13,	1.87)(14,	2.99)(15,	1.36)(16,	1.02)(17,	0.61)(18,	1.17)(19,	0.70)(20,	1.1)(21,	1.44)(22,	2.63)(23,	2.78)(24,	2.04)(25,	3.03)(26,	15.56)(27,	2.46)(28,	1.46)(29,	1.25)(30,	0.88)(31,	1.13)(32,	0.57)(33,	0.94)(34,	0.91)};
\legend{HEVC,SAP method \cite{SAP},3-tap filter}
\end{axis}
\end{tikzpicture}
\centering
\caption{Average frequency of selection for each intra mode in HEVC with different spatial prediction methods.} 
\label{fig:freq}
\end{figure*}
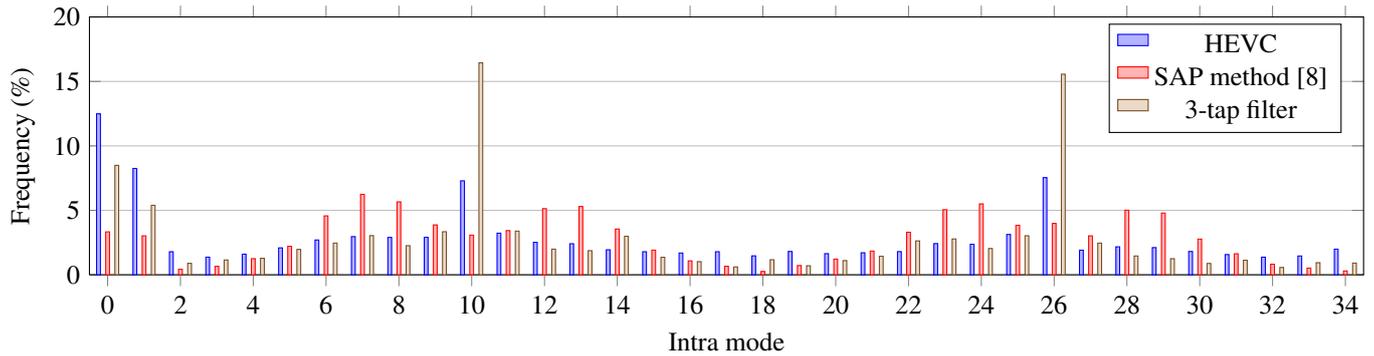

The frequency of selection for each intra mode with different prediction methods (averaged over all experiment sequences) are provided in Figure \ref{fig:freq} for HEVC, SAP and 3-tap filtering methods. It can be seen that in HEVC, planar, DC, horizontal and vertical (0,1,10,26) modes are most frequently selected. In the SAP method, planar and DC are not the most frequently selected modes anymore due to their block-based prediction. In the 3-tap filtering method, planar and DC modes are again amongst the most frequently selected modes together with horizontal and vertical modes. The frequency of selection for horizontal and vertical modes are twice as large as those of planar or DC modes.

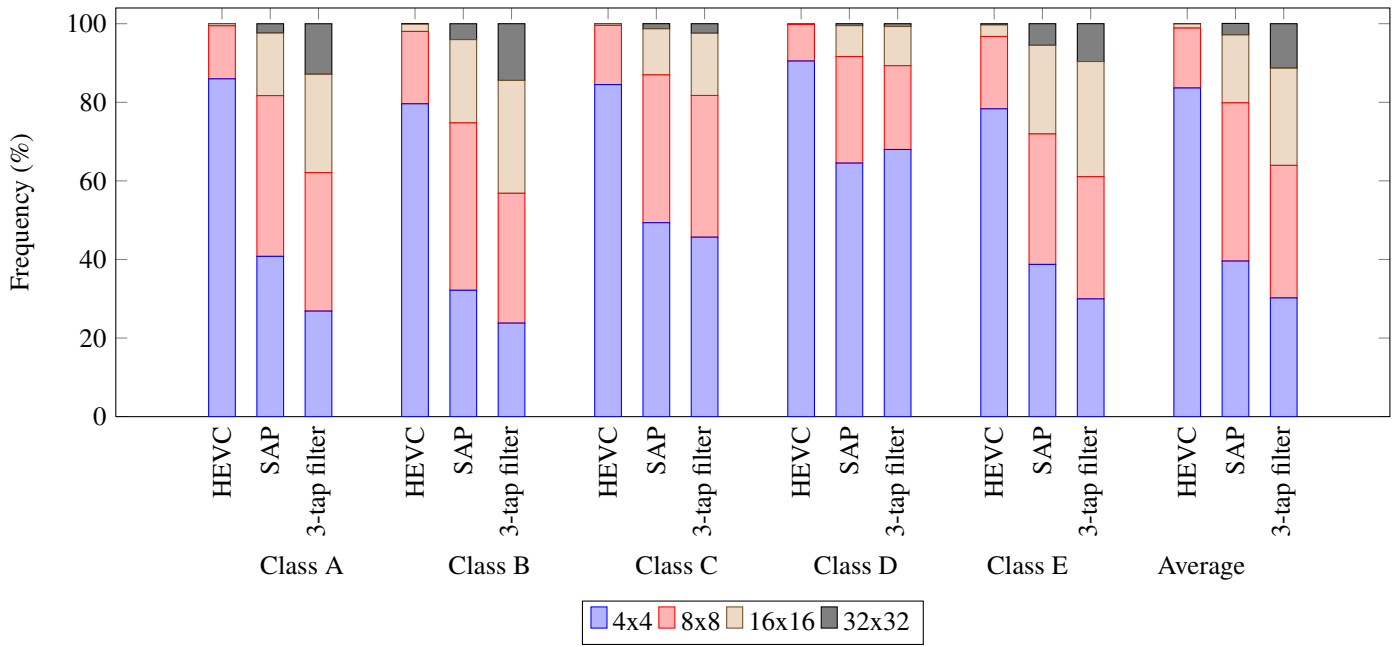
\begin{figure*}[tbh]
\centering
\begin{tikzpicture}
\begin{axis}[
width=\textwidth,
height=7cm, ymin=0, ymax=104,
    symbolic x coords={ClassAH,ClassAS,ClassA3, s1, ClassBH,ClassBS,ClassB3, s2, ClassCH,ClassCS,ClassC3, s3, ClassDH,ClassDS,ClassD3, s4, ClassEH,ClassES,ClassE3, s5, ClassTH,ClassTS,ClassT3},
     scaled ticks=false, 
    ybar stacked,
    legend style={at={(0.5,-0.45)}, anchor=north,legend columns=-1},
    scaled x ticks = false,
    x tick label style={rotate=90,anchor=east},
    x label style={at={(axis description cs:0.5,-0.22)},anchor=north},
      xlabel={ Class A \hspace{1.2cm} Class B \hspace{1.2cm} Class C \hspace{1.0cm} $~$Class D \hspace{1.0cm} Class E \hspace{1.0cm} Average},
      xtick=data,
      ylabel={Frequency (\%)},
      xticklabels={HEVC,SAP,3-tap filter, HEVC,SAP,3-tap filter,HEVC,SAP,3-tap filter ,HEVC,SAP,3-tap filter ,HEVC,SAP,3-tap filter,HEVC,SAP,3-tap filter}
    ]
\addplot+[ybar] plot coordinates {
(ClassAH,85.97) (ClassAS,40.79) (ClassA3,26.86) (ClassBH,79.62) (ClassBS,32.16) (ClassB3,23.80) (ClassCH,84.51) (ClassCS,49.36) (ClassC3,45.67) (ClassDH,90.50) (ClassDS,64.52) (ClassD3,67.99) (ClassEH,78.34) (ClassES,38.74) (ClassE3,29.97) (ClassTH,83.64) (ClassTS,39.59) (ClassT3,30.22)};
\addplot+[ybar] plot coordinates {
(ClassAH,13.50) (ClassAS,40.90) (ClassA3,35.25) (ClassBH,18.41) (ClassBS,42.60) (ClassB3,33.04) (ClassCH,15.06) (ClassCS,37.61) (ClassC3,36.06) (ClassDH,9.31) (ClassDS,27.13) (ClassD3,21.32) (ClassEH,18.41) (ClassES,33.21) (ClassE3,31.10) (ClassTH,15.25) (ClassTS,40.27) (ClassT3,33.72)};
\addplot+[ybar] plot coordinates {
(ClassAH,0.52) (ClassAS,15.92) (ClassA3,25.01) (ClassBH,1.84) (ClassBS,21.19) (ClassB3,28.72) (ClassCH,0.42) (ClassCS,11.73) (ClassC3,15.84) (ClassDH,0.19) (ClassDS,7.85) (ClassD3,9.97)  (ClassEH,2.93) (ClassES,22.55) (ClassE3,29.30) (ClassTH,1.05) (ClassTS,17.29) (ClassT3,24.74)};
\addplot+[ybar] plot coordinates {
(ClassAH,0.01) (ClassAS,2.38) (ClassA3,12.87)  (ClassBH,0.12) (ClassBS,4.04) (ClassB3,14.44) (ClassCH,0.01) (ClassCS,1.31) (ClassC3,2.43)  (ClassDH,0.00) (ClassDS,0.50) 
(ClassD3,0.72)  (ClassEH,0.31) (ClassES,5.50) (ClassE3,9.64)  (ClassTH,0.06) (ClassTS,2.92) (ClassT3,11.32)};
\legend{4x4, 8x8, 16x16, 32x32}
\end{axis}
\end{tikzpicture}
\caption{Average frequency ($\%$) of selection for each PU block-size in HEVC with different spatial prediction methods.}
\label{fig:freqPU}
\end{figure*}

The frequency of selection for each PU block-size with different prediction methods (averaged over all experiments in each class of sequences) are provided in Figure \ref{fig:freqPU}. It can be seen that in HEVC, 4x4 is by far the most frequently selected block-size with around $80\%$, and 8x8 is the next most frequently selected block-size taking almost all of the remaining percentage, and 16x16 and 32x32 block-sizes are almost not selected. This is due to the block-based prediction in HEVC, which is most accurate in the smallest 4x4 block-size. In the SAP \cite{SAP} and 3-tap filtering methods, due to the more accurate pixel-by-pixel prediction methods, 16x16 and 32x32 block-sizes are also selected with significant frequency. Notice that the frequency of selection for the 16x16 and 32x32 block-sizes are largest with the 3-tap filtering method, which can be and indication of its more effective spatial prediction performance.

\subsection{Complexity analysis}
\label{ssec:times}
This section presents a complexity analysis based on the encoding and decoding time of HEVC reference software with different spatial prediction methods. The results are normalized with respect to the HEVC encoding/decoding times and are given in Table \ref{tb:times}. These results were obtained by averaging encoding and decoding times of all frames in all experiment sequences.

Notice that, even though more computations are required to predict each block pixel in the SAP and 3-tap filtering methods than in HEVC methods, the average decoding times with SAP and 3-tap filtering methods decrease to $94.4\%$ and $87.3\%$ of the average HEVC decoding time. This is due to the bitrate reductions in the SAP and 3-tap filtering methods which allow the computationally complex entropy decoding process to finish faster. In the encoder, the faster entropy encoder is not sufficient to compensate for the increased computations of the predictions due to the repetitive rate-distortion optimized intra mode selection process, where a large number of intra modes must be computed in each block. Hence, the encoding times with SAP and 3-tap filtering methods increase to $101.8\%$ and $109.7\%$ of the average HEVC encoding time.

In addition to encoding and decoding times obtained from HEVC reference software, the following observations can be also beneficial for some implementation platforms, such as hardware implementations. First, decoder implementation is discussed. In the block-based HEVC method, all block pixels can be predicted in parallel because all depend on only the block-neighbor pixels. In the SAP method, however, there is a dependency between rows (in vertical modes) or columns (in horizontal modes) of pixels because each row/column of pixels are predicted from the reconstructed pixels in the above row/column. Hence to predict each row/column, the reconstruction (i.e. adding prediction and residual to get original pixel values) of pixels in the previous row/column must be completed first. A stronger dependency exist in the 3-tap filtering method. Each pixel is predicted from the reconstructed left and above pixels. Hence to predict each block pixel, the reconstruction of immediately previous pixel and pixels in the previous row/column must be completed first. Thus it seems that each pixel must be processed sequentially, however, an approach similar to the wavefront parallel processing \cite{HEVC} of coding tree units in HEVC can be used here also to reduce the sequential processing dependencies. In encoder implementations, all of the dependency issues can be overcome since all original pixels in a block are readily available.

\begin{table}[tbh]
\centering
\caption{Normalized average encoding and decoding times}
\label{tb:times}
\begin{tabular}{l|c|c}
\hline
                      & Encoding Time         & Decoding Time   \\ \hline \hline
HEVC method           & 100.0\%   & 100.0\%  \\ \hline                  
SAP method \cite{SAP} & 101.8\% & 94.4\% \\ \hline
3-tap filtering       & 109.7\% & 87.3\% \\ \hline
\end{tabular}
\end{table}

\section{Conclusions}
\label{sec:conc}
This paper proposed a pixel-by-pixel spatial prediction method for lossless intra coding in HEVC. The proposed method uses three neighboring pixels for prediction according to a two-dimensional correlation model. The used neighbor pixels and weights change depending on intra mode. To find the best weights for each intra mode, a two-stage offline optimization algorithm was used and a number of implementation aspects were analyzed to simplify the proposed prediction method.

Experimental results within HEVC reference software showed that when all HEVC intra modes are replaced with the proposed 3-tap filtering method, an average $11.34\%$ bitrate reduction was achieved over the default lossless intra coding in HEVC. The proposed method also decreases average decoding time by $12.7\%$ while it increases average encoding time by $9.7\%$.

\section*{Acknowledgement}
This research was supported by Grant 113E516 of T\"{u}bitak.\\



\bibliographystyle{elsarticle-num} 
\bibliography{elsarticle-template-num.bbl}



\end{document}